\documentclass[twocolumn]{aastex7}
\usepackage{graphicx}
\usepackage[version=4]{mhchem} 
\usepackage{xspace}

\newcommand{\poseidon}{\texttt{POSEIDON}\xspace}

\newcommand{\rfast}{\texttt{rfast}\xspace}
\newcommand{\Atmos}{\texttt{Atmos}\xspace}

\shorttitle{ExoTitan}
\shortauthors{Mayorga \& Lustig-Yaeger et al.}

\graphicspath{{./}{figures/}}

\begin{document}

\title{Piercing through the Haze: Characterizing Titan-like terrestrial exoplanets with JWST}

\author[0000-0002-4321-4581]{L. C. Mayorga}
\affiliation{Johns Hopkins Applied Physics Laboratory, 
Laurel, MD 20723, USA}
\email[show]{laura.mayorga@jhuapl.edu}

\author[0000-0002-0746-1980]{Jacob Lustig-Yaeger}
\affiliation{Johns Hopkins Applied Physics Laboratory, 
Laurel, MD 20723, USA}
\email{jacob.lustig-yaeger@jhuapl.edu}

\author[0000-0003-0429-9487]{Andrew P. Lincowski}
\affiliation{University of Wyoming, School of Computing, Laramie, WY 82071, USA}
\email{alincows@uwyo.edu}

\author[0000-0001-8397-3315]{Kathleen E. Mandt}
\affiliation{NASA Goddard Space Flight Center, Greenbelt, MD, USA}
\email{Kathleen.Mandt@nasa.gov}

\author[0000-0002-0766-4163]{Ryan C. Felton}
\affiliation{NASA Ames Research Center, Moffett Field, CA 94035, USA}
\affiliation{Bay Area Environmental Research Institute, Moffett Field, CA 94035, USA}
\email{ryan.c.felton@nasa.gov}

\author[0000-0002-7744-246X]{Adrienn Luspay-Kuti}
\affiliation{Johns Hopkins Applied Physics Laboratory, Laurel, MD 20723, USA}
\email{Adrienn.Luspay-Kuti@jhuapl.edu}

\author[0000-0001-7393-2368]{Kristin S. Sotzen}
\affiliation{Johns Hopkins Applied Physics Laboratory, Laurel, MD 20723, USA}
\email{Kristin.Sotzen@jhuapl.edu}

\author[0000-0002-2267-0609]{Cameron A. Gutgsell}
\affiliation{Johns Hopkins Applied Physics Laboratory, Laurel, MD 20723, USA}
\email{cam.gutgsell@jhuapl.edu}

\begin{abstract}
As we enter a new era of using JWST to characterize exoplanet atmospheres, a top priority is defining observational needs and parameters in the search for life on an exoplanet. Although some studies have begun to consider Titan-like exoplanet atmospheres, more work is needed to prepare for the potential discovery and characterization of Titan analogues with JWST and future missions. We explore the effectiveness of exoplanet atmospheric retrieval methodologies on the Cassini/VIMS-IR spectrum of Titan assuming agnostic haze properties and tholin hazes, as well as a self-consistent photochemical model. We compare our retrieved chemical abundances and haze profiles against fiducial values and Cassini/ISS-NAC UV observations and the forward modeled truths. We generalize these findings to a Titan-like exoplanet in the TRAPPIST-1 system and find that JWST-like observations have limited potential to identify features beyond methane in a Titan-like atmosphere, but may be able to identify the presence of the haze to a few hundred kilometers. The diagnostic features which would differentiate between different haze model (scattering) properties are shortward of 1\,\micron{} and further demonstrate the important role of nUV -- optical observatories in constraining the properties of exoplanet atmospheres. We also find that agnostic priors for the bulk atmosphere can lead to a degeneracy in which a spectrally active gas ($\rm CH_4$) is erroneously favored as the dominant atmospheric constituent over a spectrally inactive molecule ($\rm N_2$). 
\end{abstract}


\section{Introduction}

With the launch of JWST and looking towards HWO, a new era of exoplanet characterization has emerged with a top priority of defining the observational needs and parameters for the search for life on an exoplanet. Interpreting observations to constrain the conditions that may exist on the surface of an exoplanet can be challenging \citep{8}. Atmospheric retrievals have been used to provide observational constraints for Earth-like and Venus-like exoplanets \citep{1,2,3,Lustig-Yaeger2023Earth}. 

Ocean worlds, where a liquid water ocean is expected to exist below an ice shell \citep{9} are another example of worlds in the Solar System where life may begin to form. Titan is an ideal analog for exoplanet exploration due to not only being an ocean world but because it has a thick and extensive atmosphere with complex hydrocarbon chemistry and liquid methane \ce{CH4} and ethane \ce{C2H6} on its surface. Although life has not been detected on Titan, the surface lakes and atmospheric-surface chemical cycles remain of intense astrobiological interest as a natural lab to investigate prebiotic chemistry \citep{10,11}.

While chemical complexity has been proposed as an agnostic exoplanet biosignature \citep{walker2018exoplanet} that may be remotely assessed via disequilibrium chemistry \citep{lovelock1965physical, young2024inferring}, the moderate disequilibrium in Titan's atmosphere is driven by photochemistry \citep{krissansen2016detecting}. As a result, Titan exemplifies why geochemical and planetary system context are critically important to understanding putative biosignatures \citep{walker2018exoplanet, meadows2018exoplanet}. With an atmosphere of primarily \ce{N2} and \ce{CH4}, and a surface pressure that allows for a hydrologic-type methane cycle \citep{13}, FUV photons and energetic particles in Saturn's magnetosphere initiate complex atmospheric chemistry that leads to the formation of organic molecules, like \ce{C2H6} and \ce{HCN}, and hazes \citep{16,Robinson2014TitanWorld,18}. Despite Titan's proximity to Earth and the success of Cassini-Huygens, questions remain that illustrate the challenge for the exoplanet community in understanding \ce{N2}-\ce{CH4} dominated atmospheres, such as the escape rate of \ce{CH4} \citep{25,26,27,28}.

Titan, as the only reducing atmosphere on a small planet/moon in the Solar System, is a critical reference for the expected diversity of exoplanets. In preparation to discover and recognize exo-Titans we present an evaluation of the information inferred from the exoplanet retrieval tools \poseidon \citep{MacDonald2017HDWater, MacDonald2023POSEIDON:Spectra} and \rfast \citep{Robinson2023ExploringObservations} on the Cassini/VIMS-IR spectrum of Titan \citep{Robinson2014TitanWorld}. In particular, we evaluate the retrieved molecular abundances, temperatures, and haze properties and how the quality of these inferences is degraded in JWST-style transmission observations of TRAPPIST-1\,h. As shown in \citet{8}, TRAPPIST-1\,h has similar properties to Titan, and at present, remains the only known exoplanet that is smaller than 1.5$R_\oplus$ and 3$M_\oplus$ with the most similar insolation to Titan itself. Recent research has used the empirical Cassini/VIMS-IR Titan transmission spectrum to aid in exoplanet research in advance of more precise exoplanet measurements. This includes work by \citet{Changeat2025CloudJupiters} to simulate Titan-like hydrocarbon hazes on giant planet atmospheres and work by \citet{Niraula2025TheExoplanet} to illustrate the large number of hydrocarbons with overlapping NIR ro-vibrational bands.  

In \autoref{sec:methods}, we describe our data preparation, model setups, and retrieval setups. In \autoref{sec:results}, we show the results of our Titan and exo-Titan retrieval tests. We discuss our results and their impacts on future observations in \autoref{sec:discussion} before concluding in \autoref{sec:concl}.

\section{Methods}
\label{sec:methods}
In this work, we take a Cassini/VIMS-IR observation of Titan and an \Atmos photochemical model of Titan and attempt to retrieve the atmospheric properties for both Titan (around Saturn) and a Titan-like atmosphere on TRAPPIST-1\,h. In the case of retrieving on actual Titan, we use both the Cassini observation and an \Atmos photochemical model as described in this section to validate our retrieval an modeling methodologies and examine the inferences we can make at such observational precisions. In the case of retrieving on a Titan-like TRAPPIST-1\,h we scale up the Cassini observation as a test of detectability, but have a self-consistent \Atmos model to explore the real chemical and haze properties of such a planet. In this section we describe the preparation of our datasets and modeling frameworks.

\subsection{Cassini Sample}
Rather than averaging the Cassini/VIMS-IR Titan transmission spectra like \citet{Niraula2025TheExoplanet}, we follow \citet{Robinson2023ExploringObservations} and examine the 2011 September observation from 27\degr{}N. This dataset was selected due to the overlapping temporal and spatial coverage with the Cassini/ISS-NAC UV3 dataset in \citet{Seignovert2021HazeMission}. In \citet{Seignovert2021HazeMission} latitudinal haze extinction profiles are retrieved to examine the thin global detached haze layer above the main global haze layer. In the observation N1694250019 taken also in 2011 September, this detached layer has dropped and disappeared and, thus, we avoid confounding our interpretations. Haze extinction is shown to increase from 10$^{-7.8}$ near 550\,km to 10$^{-6}$ near 350\,km while the occultation observations probe heights of 100--350\,km, much deeper in the atmosphere than the detached haze layer.

We scale the Cassini/VIMS-IR Titan transmission spectrum to TRAPPIST-1\,h to assess it in the context of exoplanet observations with JWST. To do this while preserving physical differences in the zero bond albedo equilibrium temperature ($\rm T_{eq}$ of ${\sim}90$ vs ${\sim}169$ K for Titan and TRAPPIST-1h, respectively) and surface gravity ($g$ of $\rm0.138g$ vs ${\sim}\rm0.56g$ for Titan and TRAPPIST-1h, respectively) of the two bodies, we scale the spectrum (expressed in units of km of effective absorbing radius) by a ratio of scale heights for the two bodies ($H_{T1h}/H_{Titan} \approx 0.467$). This results in an expected absorbing radius of TRAPPIST-1\,h's atmosphere assuming it possesses Titan's atmosphere. We subtract the wavelength-dependent mean from the spectrum and then add the measured radius of TRAPPIST-1\,h to arrive at the wavelength dependent planet radius that is consistent with measured transit light curves. Finally, we square the planet radius and divide by the square of the TRAPPIST-1 stellar radius to obtain our estimated transmission spectrum of TRAPPIST-1\,h with Titan's atmosphere. We note that while this method does approximately account for different mass, radius, and instellation values between TRAPPIST-1\,h and Titan, it makes no attempt to adjust the atmospheric chemistry in response to these changes. 

\begin{figure}
    \includegraphics[width=\linewidth]{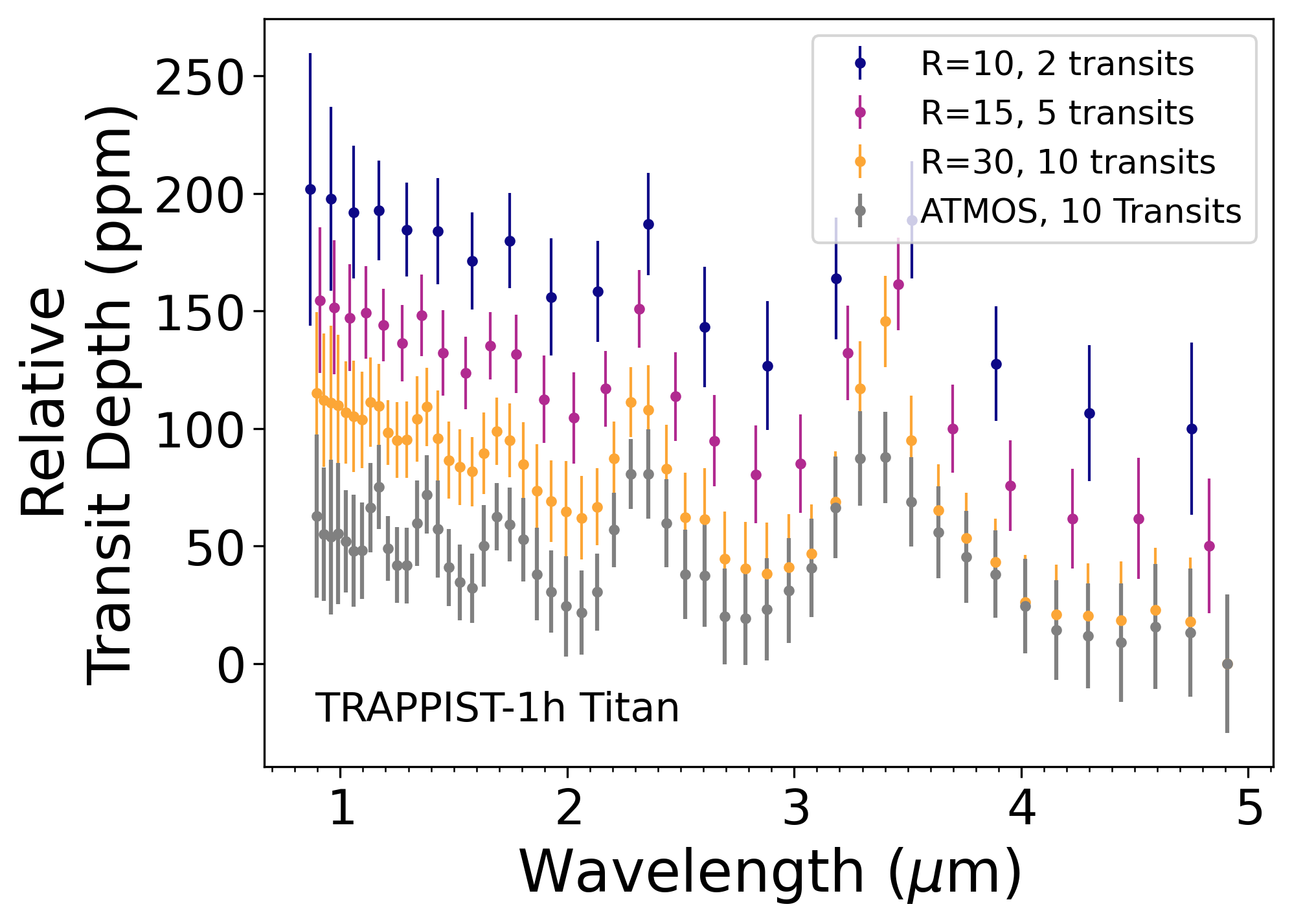}
    \caption{Spectrum of TRAPPIST-1\,h assuming it is a Titan-like atmosphere as observed by JWST with 2, 5, or 10 transits as derived by scaling the Cassini spectrum of Titan (colors) or from the \Atmos self-consistent model (gray). Cassini spectra are offset for clarity while the \Atmos model is aligned with the Cassini 10 transit spectrum at the reddest wavelength. The scaled data and self-consistent simulation differ most in the slope produced by hazes in the atmosphere.}
    \label{fig:scaled}
\end{figure}

We then use \texttt{PandExo} \citep[v3.0;][]{Batalha2017} to determine the errors on the spectrum for a single transit observation of TRAPPIST-1\,h. We then coadd and bin the simulated spectra to generate model-independent synthetic transmission spectra that could result from a 2, 5, or 10 transit observational campaign at corresponding spectral resolutions of R=10, 15, or 30, respectively. The resulting JWST-style spectra are shown in \autoref{fig:scaled} and are not too dissimilar from the self-consistent model with \Atmos (see \autoref{sec:atmos}). Note that random jitter was not added. These spectra serve as a benchmark for the detectability of Titan-like atmospheres, but are not self-consistent.

\subsection{\Atmos model} \label{sec:atmos}
To self-consistently model gas profiles and haze production with neutral chemistry for exo-Titan atmospheres, we use the 1D photochemical-kinetics model component of \Atmos, which originates from \citet{Kasting1979}, with major updates completed by \citet{Zahnle2006}, \citet{Arney2016}, and \citet{Lincowski2018}.
A separate paper will detail the updates made for the Titan components. In short, these include condensation of hydrocarbon species (\ce{H2O}, \ce{CO2}, \ce{CH4}, \ce{C2H2}, \ce{C2H4}, \ce{C2H6}, \ce{C3H8}, HCN, \ce{HC3N}, \ce{NH3}, \ce{N2H4}, \ce{CH3NH2}, \ce{CH2NH}, \ce{CH2CCH2}, \ce{CH3C2H}, and \ce{C3H6}, using the saturation vapor pressure formulas tabulated by \citealt{Krasnopolsky2009}) and includes a haze formation scheme, which has been adjusted from \citet{Arney2016}. In particular, we employ three final haze-generation pathways, after which the precursor molecules (\ce{C4H2}, \ce{C5H4}, and \ce{C2H3N2}) are presumed to form haze: 
\begin{align}
    \ce{C2H} + \ce{CH3C2H} & \rightarrow \ce{CH3} + \ce{C4H2} \label{rxn:haze1}\\
    \ce{C2H} + \ce{CH2CCH2} & \rightarrow \ce{H} + \ce{C5H4} \\
    \ce{H2CN} + \ce{HCN} & \rightarrow \ce{C2H3N2}.
\end{align}
In practice, (\ref{rxn:haze1}) dominates the total haze production, and so \ce{C4H2} is the species we use to model the haze aerosols, as in \citet{Arney2016}.

The formation scheme for aerosol size distributions is that presented in \citet{Pavlov2001a}, and used more recently in \citet{Arney2016} and \citet{Meadows2023}. The formation scheme is based on sedimentation and formation timescales. This model produces layer-by-layer monodisperse aerosols, which are treated radiatively in the photochemical model, potentially self-shielding. Optical properties for these tholin haze particles were calculated using the optical constants of \citet{Khare1984OpticalFrequencies}.

To provide simulated ``ground truth'' for our retrieval tests, we model the resultant spectra of our Titan and ExoTitan atmospheres using the transit mode of the Spectral Mapping Atmospheric Radiative Transfer (SMART) model \citep{Meadows1996,Crisp1997,Robinson2017a}, which requires calculated line-by-line absorption coefficients using LBLABC \citep{Meadows1996}. Line data for visible--IR transitions are from HITRAN2016 \citep{HITRAN2016} and UV absorption cross-sections are described in \citet{Lincowski2018}, appendix~B. Instead of using monodisperse aerosols in our spectral model, we convert the monodisperse radii into spherical or fractal hazes (depending on particle effective radii) following the procedures in \citet{Arney2016}, which includes realistic particle distributions generated for every atmospheric layer that conserves mass from the photochemical model. The transit spectra produced by SMART are output at 1~cm$^{-1}$ resolution and include the following gas species: \ce{H2O}, \ce{H2}, CO, \ce{C2H2}, \ce{C2H4}, \ce{C2H6}, \ce{C3H8}, HCN, \ce{HC3N}, \ce{NH3}, and \ce{N2}. 

We test and calibrate our Titan model template against the Cassini observation of Titan transiting the Sun presented by \citet{Robinson2014TitanWorld}. Our nominal Titan model uses the standard, recommended version of the pre-Huygens engineering model \citep{Yelle1997}, which is in excellent agreement with Huygens data in the lower atmosphere but does deviate in the stratosphere and mesosphere, which is attributed to dynamical wave phenomena \citep{Fulchignoni2005}. We specify an eddy diffusion profile to produce a haze that closely matches the Cassini transit spectrum. In our current model, this results in less optimal profiles for the primary hydrocarbons (\ce{C2H2}, \ce{C2H4}, and \ce{C2H6}), which are nonetheless within the range of observations of these species in Titan's atmosphere \citep[e.g.][]{Vinatier2010,Koskinen2011,Nixon2013}, except for \ce{C3H8}, which is a factor of $\sim2$ higher. We do note \ce{C2H2} in particular has an enhanced feature in our spectrum at 3~$\mu$m. However, there is a larger and broader discrepancy there caused by excessive tholin haze absorption. Observations of Titan's haze have not yet been reconciled with laboratory studies (e.g.  Khare's optical properties), which suggests the hazes are more layered and/or chemically inhomogeneous \citep[e.g.][]{He2022}. The spectrum is very similar to Titan's observed spectrum with some slope difference at shorter wavelengths and a vertical offset due to the simulated spectrum being used for the retrieval tests was based on a 1~bar Titan model atmosphere rather than Titan's true 1.5~bar surface. 

To model TRAPPIST-1\,h as a potential exoTitan, we employ the planetary parameters from \citet{Agol2021}, i.e. radius, gravity, semi-major axis, and impact parameter. We use an average of the stellar spectrum for TRAPPIST-1 produced by \citet{Peacock2019} and presented by \citet{Meadows2023} for both the photochemical model and SMART. As our temperature profile, we shift Titan's nominal profile to a surface of 150~K (approximately TRAPPIST-1 h's equilibrium temperature) and re-grid the temperature vs. altitude for the surface gravity of TRAPPIST-1~h. We follow the same modeling procedures described above for Titan, including spectral resolution and included gases.
The model spectrum with the appropriate noise for a 10 transit observation is shown in \autoref{fig:scaled}.

\subsection{Retrievals}
To facilitate the comparisons between \poseidon and \rfast we make agnostic and simplistic assumptions about the ``planet's" atmosphere. Our primary assumptions included an isothermal temperature-pressure profile \citep[in][they adopted a three parameter temperature model]{Robinson2023ExploringObservations} and fit for the radius, the respective haze parameters and the following species: \ce{CO}, \ce{CO2}, \ce{CH4}, \ce{C2H2}, \ce{C2H4}, \ce{C2H6}, \ce{C3H8}, and \ce{HCN}. As demonstrated by \citet{Niraula2025TheExoplanet}, hydrocarbons have many lookalikes and degenerate features particularly from 3--3.5\,\micron{} due to the nature of C-H bonds. 

The \rfast \citep{Robinson2023ExploringObservations} retrieval model is an open source atmospheric retrieval package capable of generating transmission, reflection, and emission forward models and was developed to support direct imaging mission concepts. With \rfast we use 90 layers spaced logarithmically between 10$^{-3}$ and 10$^{6}$\,Pa with the radius defined at the bottom, 10$^{6}$\,Pa. \rfast also includes Rayleigh scattering and collisionally induced absorption (CIA) from \ce{N2} and \ce{H2}.

The \poseidon \citep[v1.2;][]{MacDonald2017HDWater, MacDonald2023POSEIDON:Spectra, Mullens2024} retrieval model is also used to fit the Titan transmission spectrum in the NIR. Our \poseidon retrieval models use 200 layer atmospheres (log-sampled from 100 to $10^{-7}$ bar). In addition to the aforementioned gases, \poseidon includes CIA from \citet{Karman2019} for \ce{N2}-\ce{N2}, \ce{N2}-\ce{H2O}, \ce{CO2}-\ce{CO2}, and \ce{CO2}-\ce{CH4} \citep[see][for more details on the \poseidon opacity database]{MacDonald2022}. 

\begin{figure}[t]
    \includegraphics[width=\linewidth]{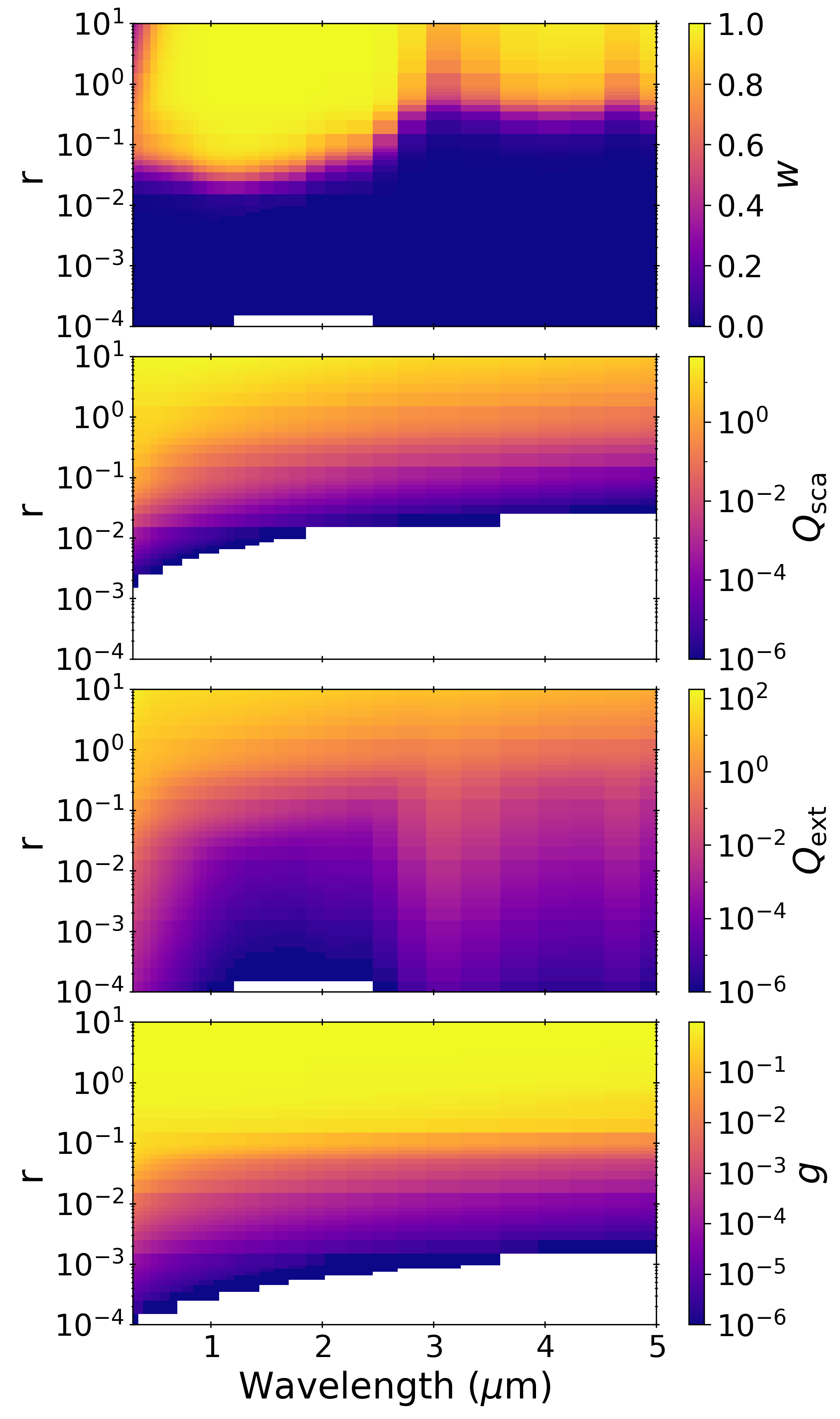}
    \caption{Khare tholin properties \citep{Khare1984OpticalFrequencies}. Where $w = Q_\mathrm{sca}/Q_\mathrm{ext}$. For any given size, we interpolate in radius to produce the particle parameters.}
    \label{fig:khare}
\end{figure}

We take two approaches to modeling the haze in Titan's atmosphere. In \rfast, we developed a model for the physical extent of the haze layer based off of its minimum and maximum optical depth, $\tau_{min}$ and $\tau_{max}$, and the haze top and base pressure, $p_{H,min}$ and $p_{H,max}$. We then populate this haze layer evenly with haze aggregates of a given size or sizes using the optical properties of the tholins in \citet{Khare1984OpticalFrequencies}, which are shown in \autoref{fig:khare}. We interpolate in wavelength and aggregate radius space, $s_{small}$ to input the scattering efficiency $Q_{scat}$, and the asymmetry parameter, $g$, and the single scattering albedo, $w$, which is the ratio between $Q_{scat}$ and the extinction efficiency, $Q_{ext}$. In the case of two sizes of aggregates, we model the population with a fraction of the smaller size, $f_{small}$, of each size. In \poseidon, we instead take the more agnostic exoplanet approach of modeling the haze as an enhancement to the Rayleigh scattering slope \citep{LecavelierDesEtangs2008Rayleigh189733b, MacDonald2017HDWater}. In this scenario, the extinction is given by
\begin{align}
    \label{eqn:powerlaw}
    \kappa_\mathrm{cloud}(r) =
    \begin{cases}
        a \sigma_0(\lambda/\lambda_0)^\gamma  &(P<P_\mathrm{cloud})\\
        \infty & (P \geq P_\mathrm{cloud}),
    \end{cases}
\end{align}
where $a$ is the Rayleigh-enhancement factor, $\sigma_0$ is the \ce{H2}-Rayleigh scattering cross section at the reference wavelength, $\lambda_0$, $\gamma$ is the scattering slope, and r is the radial coordinate \citep[as shown in Figure 1 of ][]{MacDonald2017HDWater}.

Each model fits for its own cloud parameters, the planet radius, the surface temperature, and the volume mixing ratio of each gas. \poseidon directly fits for \ce{N2} using center-log-ratio (CLR) priors (see below) while \rfast uses \ce{N2} as the filler gas. \rfast runs used logarithmic priors on the mixing ratios, between 10$^{-12}$--1, and logarithmic priors on the haze parameters of the size or size fraction, and the pressure span, and linear for the optical depths, radius, and temperature. Molecular opacities are generally derived from
the HITRAN database \citep{HITRAN2020} using the Line-By-
Line ABSorption Coefﬁcients tool \citep[LBLABC, ][]{Meadows1996}.

For \poseidon, we adopt a common set of free parameters with uniform priors. These are:  the reference radius at 10 bars ($\mathrm{R}_{\mathrm{p, ref}} \sim \mathcal{U}(0.05055, 0.07414) ~ \mathrm{R_J}$), the isothermal atmospheric temperature ($\mathrm{T} \sim \mathcal{U}(0, 400) ~ \mathrm{K}$), the pressure of an optically thick gray cloud deck ($\log \, \mathrm{P}_{\mathrm{cloud}} \sim \mathcal{U}(-6, 2) ~ \mathrm{bar}$), a power law haze (see \autoref{eqn:powerlaw}) with Rayleigh enhancement factor relative to \ce{H2} ($\log \, \mathrm{a} \sim \mathcal{U}(-4, 8)$) and power law exponent ($\gamma \sim \mathcal{U}(-20, 2)$), and volume mixing ratios ($\log \, \mathrm{X} \sim \mathcal{U}(-12.0, 0.0)$) for the aforementioned molecules \citep{HITRAN2016_XSC, HITRAN2020}. We added cross sections from HITRAN \citep{HITRAN2016_XSC} for \ce{C3H8} to \poseidon using the \texttt{Cthulhu} Python package \citep{Agrawal2024}. With our \poseidon retrievals, we elect to use CLR transformed priors for gas volume mixing ratios (VMRs) \citep{Benneke2012AtmosphericSpectroscopy}, which implicitly satisfies the requirement that the VMRs of all species sum to unity, without imposing a bias that favors high abundances of the bulk gas that is assumed \citep[e.g.,][]{gomez2023venus, cadieux2024transmission}. 

Additional differences between the two retrievals are that \poseidon uses nested sampling via \texttt{MultiNest}/\texttt{PyMultiNest} \citep{feroz_multinest_2009, buchner_x-ray_2014} (with 1000 live points) while \rfast uses Markov Chain Monte Carlo (MCMC) via \texttt{emcee}, hence they have different convergence methodologies and criteria. Additionally, \poseidon fits for an error inflation term, $b$ \poseidon \citep[See Equation 3 from][]{line2015uniform}, that accounts for underestimated measurement errors and/or model deficiencies by inflating the errors used by the retrieval by $\sigma_{\rm infl} = \sqrt{10^b}$, which adds in quadrature with the standard errors. 

\section{Results}
\label{sec:results}

\subsection{The Baseline: Titan}
\begin{figure}[b]
    \includegraphics[width=\linewidth]{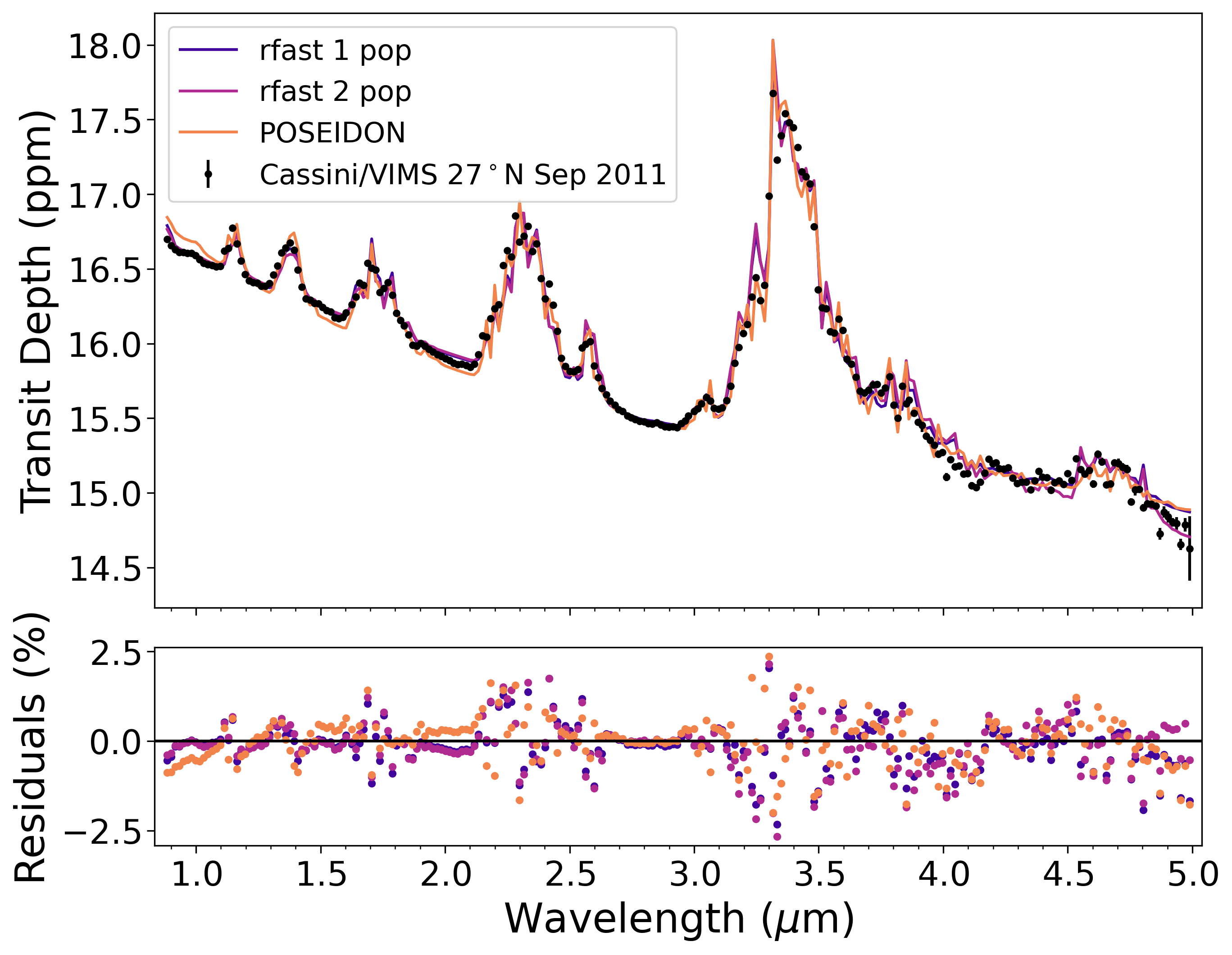}
    \caption{Best fit models from all three retrieval set ups. Retrievals mainly show percent-level discrepancies around the methane features.}
    \label{fig:titanbest}
\end{figure}

We conducted two tests with \rfast where we varied the haze population that composed the layer. In the `1 pop' case, we fit for the size, $s$, of a single size aggregate population. In the `2 pop' case, we assume the population is made up of aggregates of two fixed sizes, 0.07 and 6.0\,\micron{}, which span the change in optical property behavior (see \autoref{fig:khare}), and fit for the ratio of small to large particles, $f$, that compose the haze layer. The results of both \rfast retrievals and the \poseidon retrieval are shown in \autoref{fig:titanbest}.

\begin{figure*}[t]
    \includegraphics[width=\linewidth]{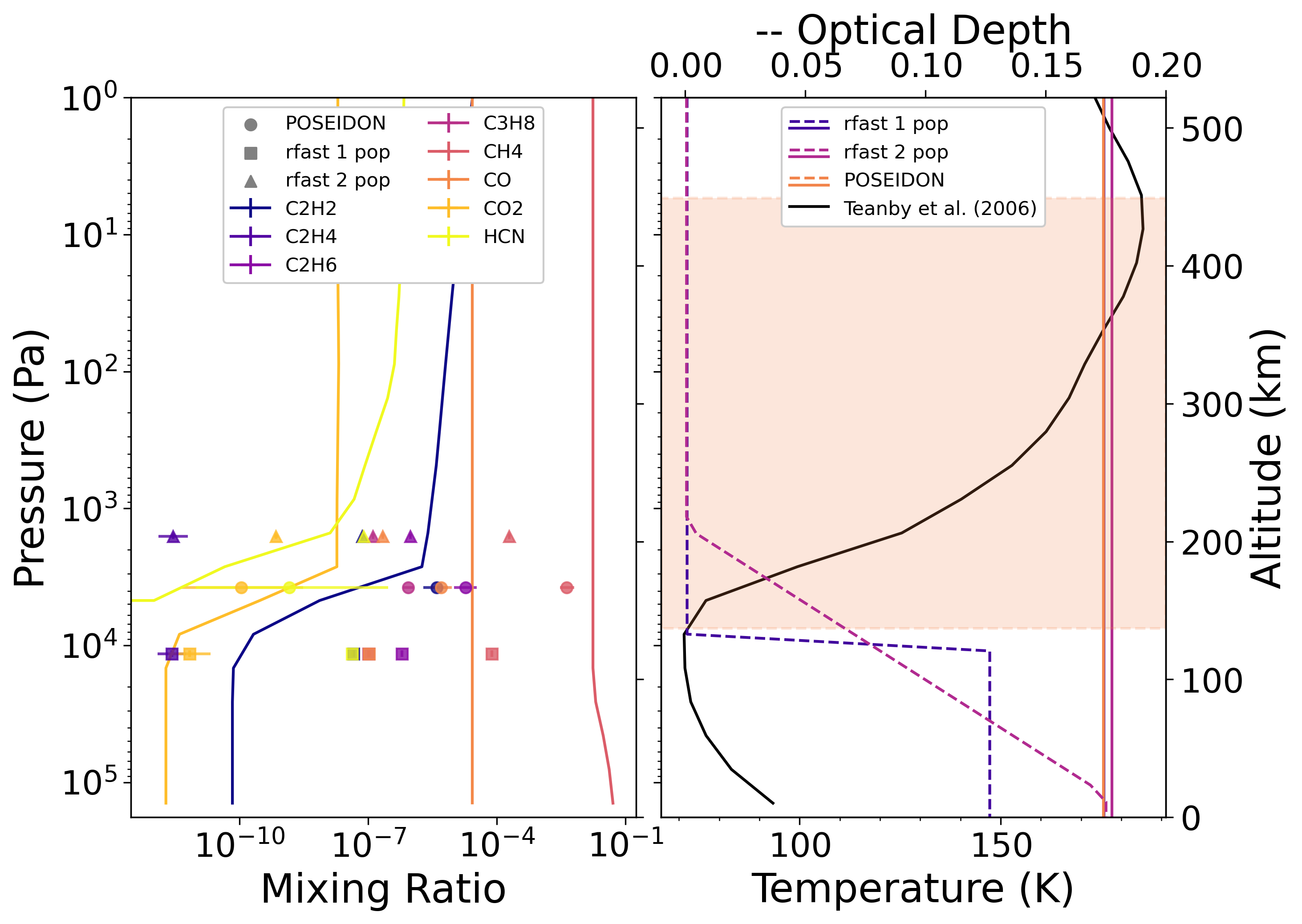}
    \caption{Retrieved temperature, mixing ratios, and haze parameters from all three retrieval set ups. Left: Retrieved volume mixing ratios for all three retrieval cases shown at the approximate depth of the haze layer. Right: Haze opacity in dashed lines and temperature in solid lines as a function of pressure and altitude. Where available we show the \citet{Teanby2006} mixing ratios and temperature profile. Despite varying assumptions and haze schemes, all retrievals indicate a haze layer in approximately the same pressure location. Note that the \poseidon{} haze location is inferred from the pressure contribution functions and is shown as a shaded region.}
    \label{fig:titanhaze}
\end{figure*}

All three retrievals reproduce Titan's spectrum as observed by Cassini to within the few percent level. The largest deviations occur within the methane features. In \autoref{fig:titanhaze}, we show the atmospheric components that were inferred from each test case. \poseidon's use of a haze slope requires additional care when determining the location of the haze since no actual haze layer, as defined by the inclusion of particles of particular optical properties, is included. The $P_\mathrm{cloud}$ is mostly sensitive to the deepest possible pressure probed, which corresponds to the longest wavelength included in the retrieval because a gray cloud, if present, must exist below (at higher pressures than) the haze scattering slope. To determine the extent of the haze, in such a paradigm, we computed wavelength-dependent pressure contribution functions and identified the pressure range that contributes the bulk of the transmission spectrum continuum. We denote this pressure range as the ``haze layer" in \autoref{fig:titanhaze} for comparison with the rfast and \Atmos results models, which both have a defined. 

Despite the variations in haze treatment, all three cases show a haze layer above roughly 100\,km in altitude. Note that the hazes in the Cassini/ISS-NAC UV3 dataset \citet{Seignovert2021HazeMission} suggest the haze extinction begins to increase starting as high as roughly 450\,km. The mixing ratios retrieved are evenly mixed in pressure, but for clarity we show them as a point near onset of haze opacity for \rfast cases and near the base of the haze contribution for \poseidon as representative of where the atmosphere is being probed. There are some variations between the retrieved cases as to the volume mixing ratios of any particular species but all three test cases show a methane dominated atmosphere with trace amounts of other carbon-bearing species. The temperatures retrieved in each case are to within a few Kelvin. Seeing little distinction between the two \rfast haze models in this wavelength range (see \autoref{sec:discussion}), we move forward in the rest of this work with a single-sized population of haze aggregates for \rfast retrievals.

\subsection{\Atmos Titan}
\begin{figure}
    \includegraphics[width=\linewidth]{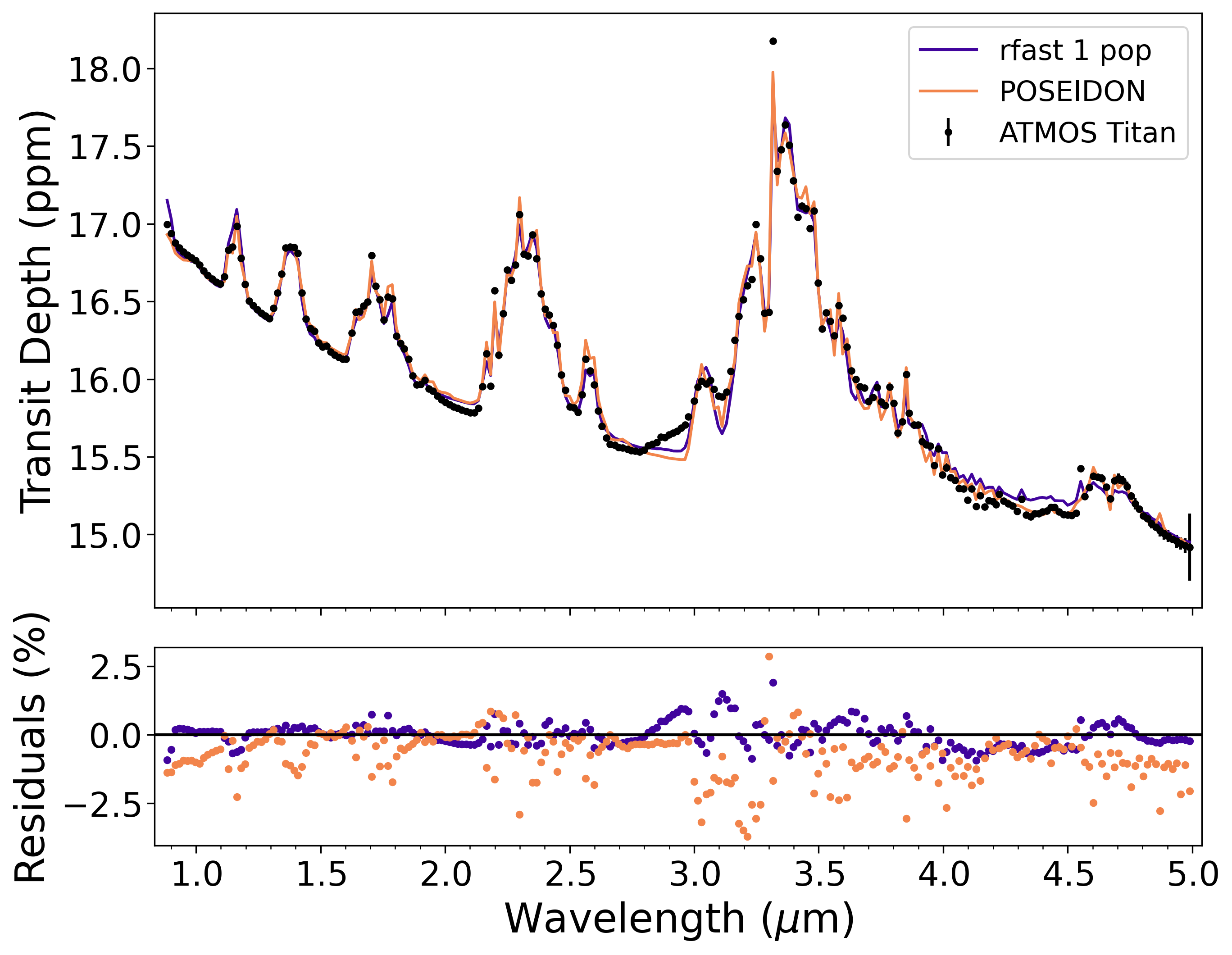}
    \caption{Best fit retrievals on the \Atmos Titan model. Retrievals mainly show percent-level discrepancies around the methane features.}
    \label{fig:atitanbest}
\end{figure}

We take the \Atmos model of a Titan-like atmosphere and interpolate it onto the Cassini observations wavelength grid and apply the Cassini uncertainties in order to produce a Cassini-like observation.
We conducted retrievals on the \Atmos Titan spectrum agnostic of the true vertical chemical, cloud, and temperature profiles and fitted for the same set of gases as in our Titan validation retrievals. \autoref{fig:atitanbest} shows the \Atmos Titan model, which is very similar to the Cassini observed spectrum, and the best fit results from both \rfast and \poseidon retrieval frameworks. The residuals are generally improved for \rfast with the same \% magnitude deviation in the 3.2\,\micron{} region.

\begin{figure*}
    \includegraphics[width=\linewidth]{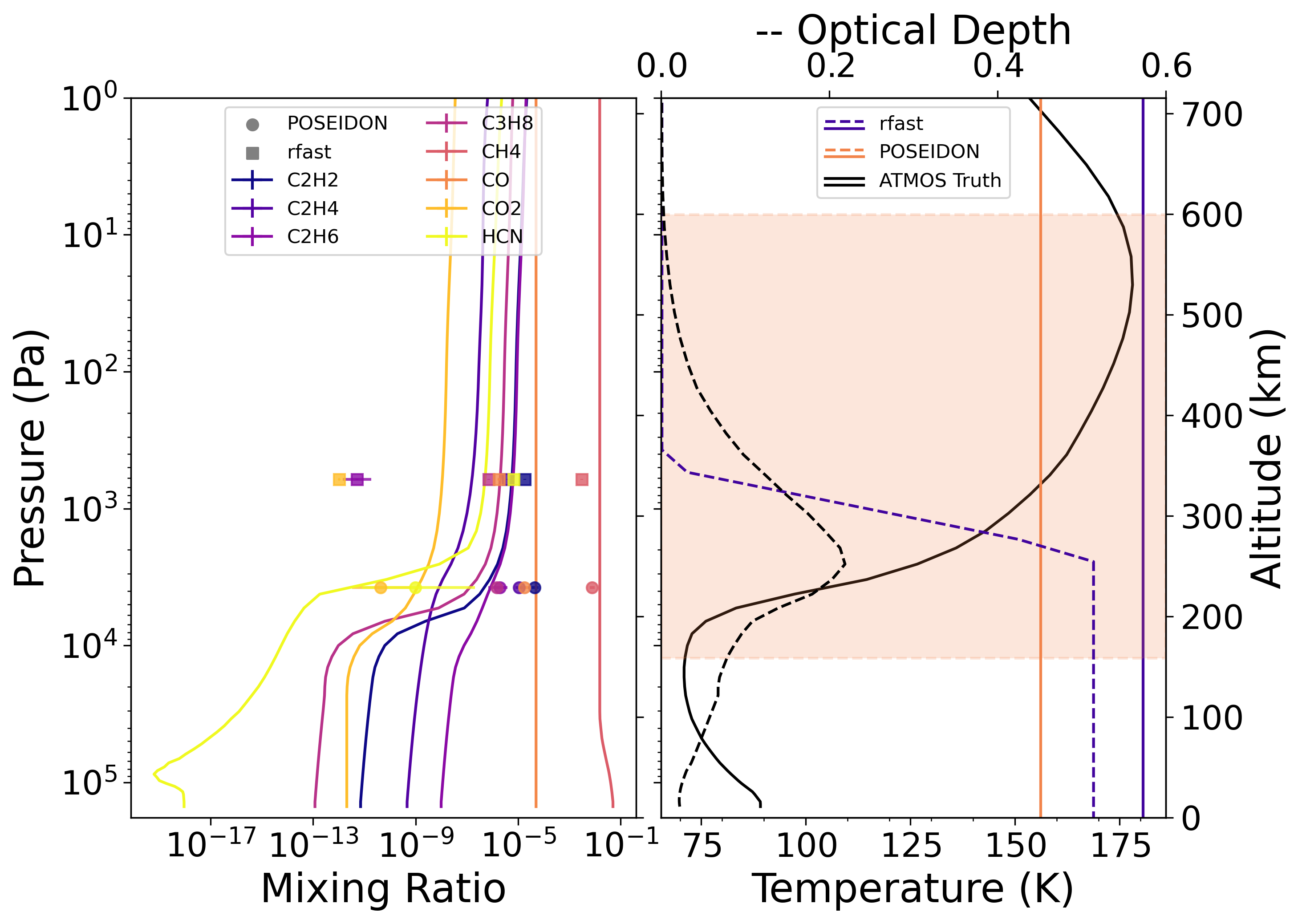}
    \caption{Same as \autoref{fig:titanhaze}, but for the \Atmos Titan model. The hazes from both retrievals are approximately at the correct altitude. More transits, i.e. high precisions, are needed to constrain much more than the methane abundance, the radius, and the pressure onset of the haze, which is generally larger particles. Note that the \poseidon{} haze location is inferred from the pressure contribution functions.}
    \label{fig:atitanhaze}
\end{figure*}

In \autoref{fig:atitanhaze}, we show the atmospheric components that were inferred by each retrieval compared to the true values in the forward model. Determining the haze layer for \poseidon in the same method as before, we see that the approximate location of the haze in both retrieval frameworks is consistent with the truth. The \rfast haze opacity peaks very close to the true pressure. Both retrievals suggest a methane dominated atmosphere, but deviate in the precise abundances of the hydrocarbons. Unlike in the Cassini case, where the retrieved temperatures were similar, the temperature retrieved by \poseidon in the \Atmos case is cooler than that of \rfast.

\subsection{Extrapolating to Exo-Titans}
\label{sec:results:exo}

\begin{figure}
    \includegraphics[width=\linewidth]{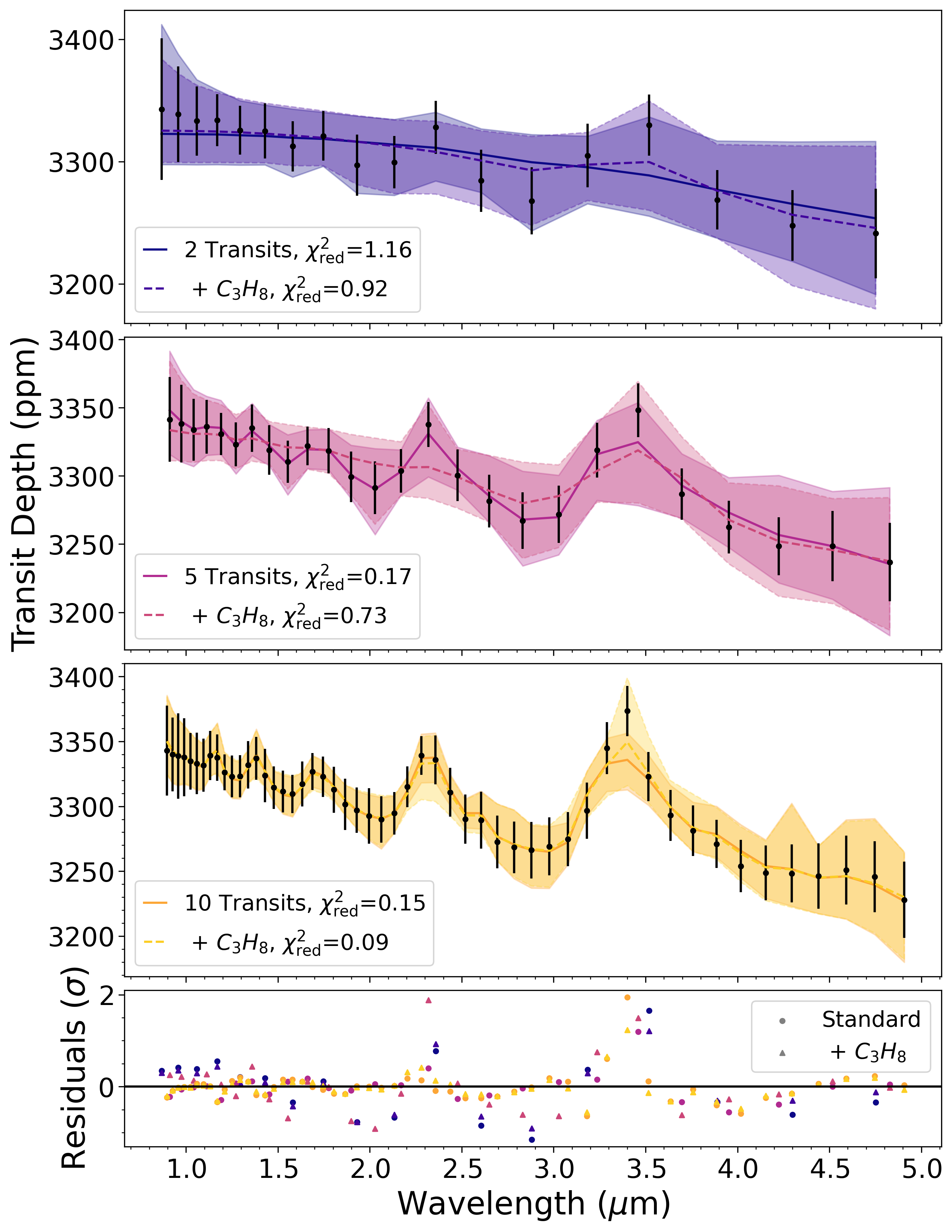}
    \caption{Best fit \rfast models and error envelopes for each scaled Cassini Titan transit fidelity scenario with and without \ce{C3H8}. Like in the Titan retrievals the discrepancies are greatest around the methane features and at significances that are suggestive and not generally improved by the addition of \ce{C3H8}.}
    \label{fig:exotitanbest}
\end{figure}

In the exo-Titan case, we used the single population \rfast retrieval set up and fit for only the common gases \ce{H2O}, \ce{CO2}, \ce{CH4}, and \ce{CO} to mimic a semi-agnostic first retrieval of a rocky exoplanet. 
We note that with JWST quality errors and no jitter, the 2 transit scenario is consistent with a featureless, but sloped spectrum, while the 5 and 10 transit scenarios, begin to favor features over a featureless null hypothesis model. 
With the comparatively large error bars of the exo-Titan observations as compared to the Cassini observations the retrievals match the spectrum quite well in all three cases. The largest deviations are seen near 2.3\,\micron{} and 3.4\,\micron{}, however these are less than 2$\sigma$ significant. While such discrepancies were seen even when fitting the Cassini observations, an astronomer seeing an exo-Titan for the first time might be tempted to search for additional species that may account for these features.
We added \ce{C3H8} to account for the C--H stretch bond that is particularly important at these discrepant wavelengths. The inclusion of propane results in marginal changes to the goodness of fit (-4$<\Delta$BIC$<$6) and is not always favored. The results of retrievals done with and without \ce{C3H8} are shown in \autoref{fig:exotitanbest}. 

\begin{figure*}
    \includegraphics[width=\linewidth]{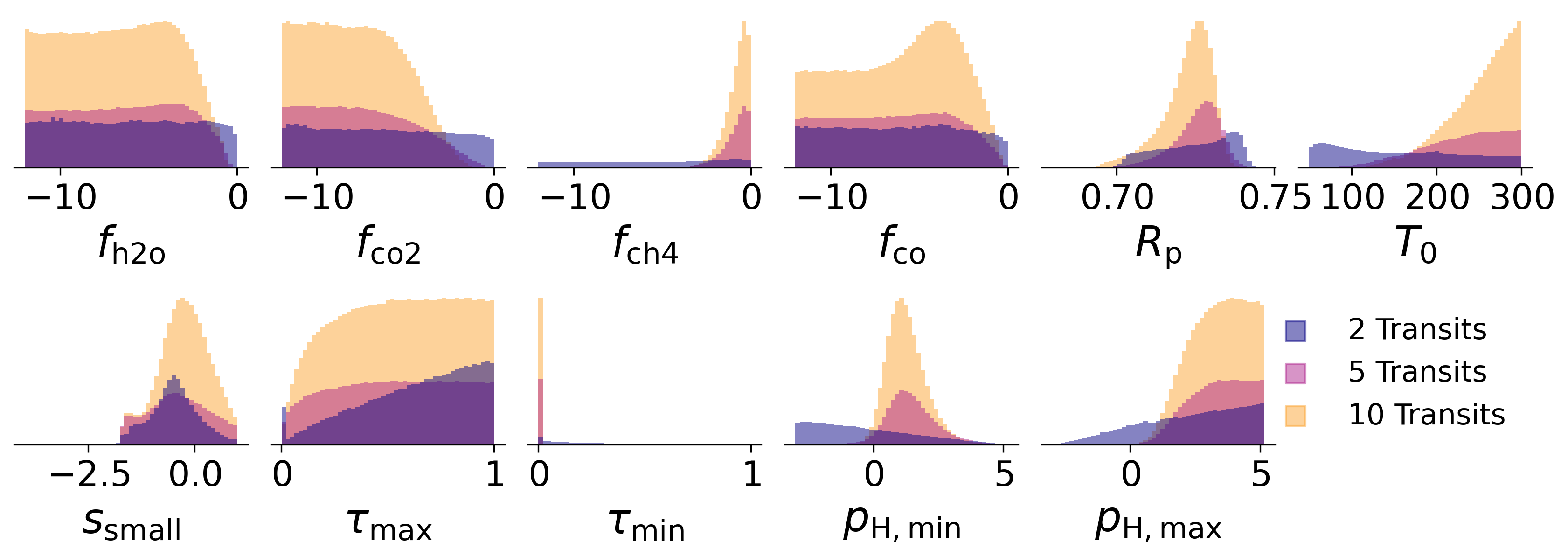}
    \caption{A comparison of the \rfast 1D posteriors for each retrieved parameter for each scaled Cassini Titan transit fidelity scenarios without \ce{C3H8}. In the 2 transit case, no real constraints are placed on the mixing ratios of the species, but in the 5 and 10 transit cases methane is readily constrained with suggestive limits placed on the other species.}
    \label{fig:posteriors}
\end{figure*}

We show the 1D posteriors for each retrieved parameter in \autoref{fig:posteriors} for each transit scenario. Generally increased precision leads to tighter constraints on the parameter (such as in the abundance of methane, $f_{\ce{ch4}}$), or a shift in the bias of the parameters distribution (such as in the radius, $R_p$). Both of these effects are notable in retrieving the temperature, which went from cold to too hot, and the start of the haze layer, $p_{H,min}$, which was very high in the atmosphere to deeper in the atmosphere. However, none of these transit scenarios have sufficient precision to motivate or identify the search for additional species nor are any scenarios particularly constraining for, specifically, \ce{H2O}, \ce{CO2} and to some extent \ce{CO}. In the 2 transit case, the volume mixing ratios of the four species are not constrained, but in the 5 and 10 transit cases, a \ce{CH4}-dominated atmosphere is readily interpretable. The detection of the methane is the key ingredient to achieving a constraint in the temperature and haze location.

\begin{figure*}
    \includegraphics[width=\linewidth]{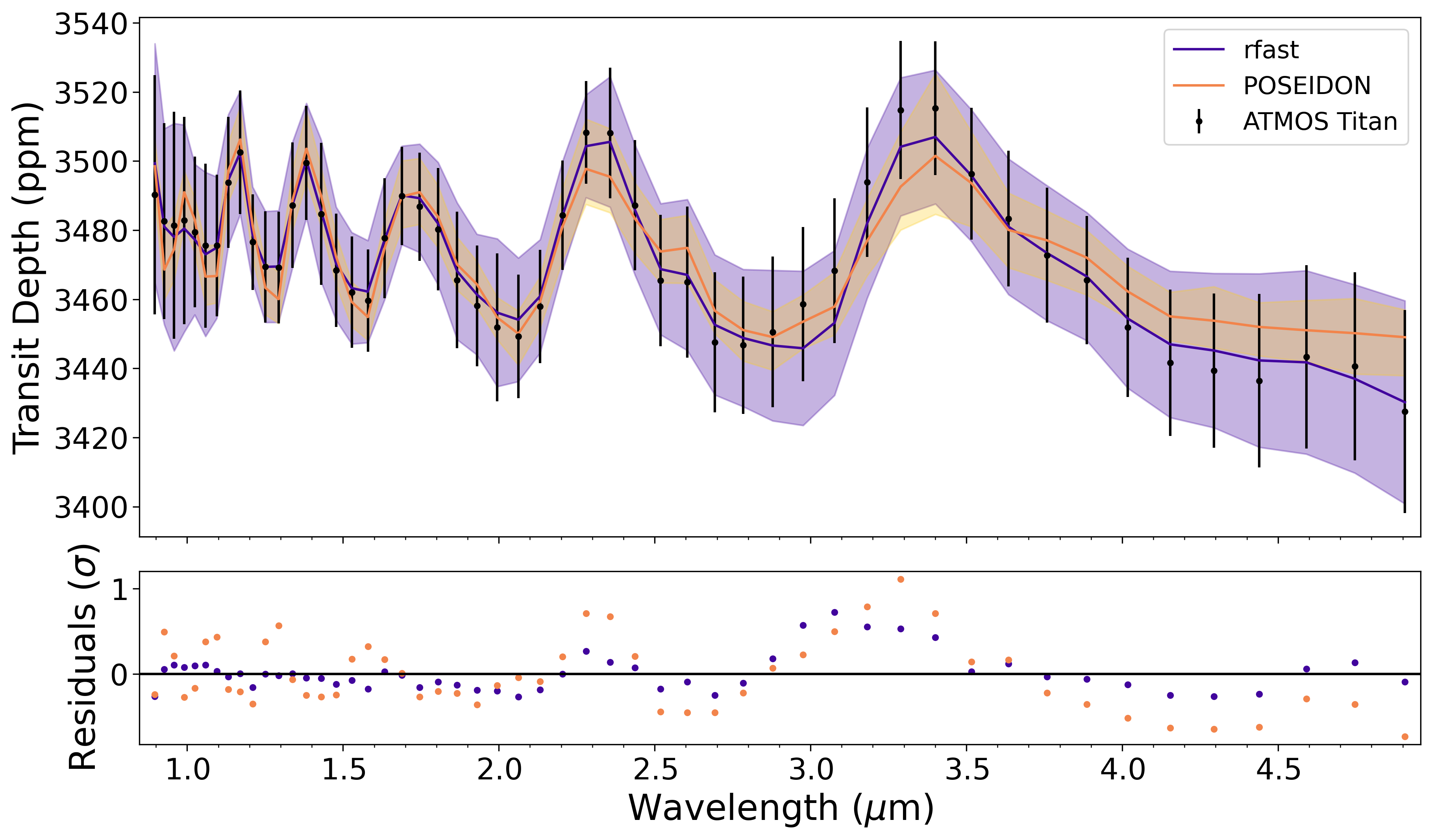}
    \caption{Best fit retrievals on the \Atmos exo-Titan model. Retrievals are consistent and do not show significant discrepancies but are suggestive around the methane features.}
    \label{fig:aexobest}
\end{figure*}

For the \Atmos self-consistent exo-Titan case, we show the results from the 10 transit scenario from both \rfast and \poseidon in \autoref{fig:aexobest}. The retrievals are consistent with each other and the residuals are small and again confined to the methane features.

\begin{figure*}
    \includegraphics[width=\linewidth]{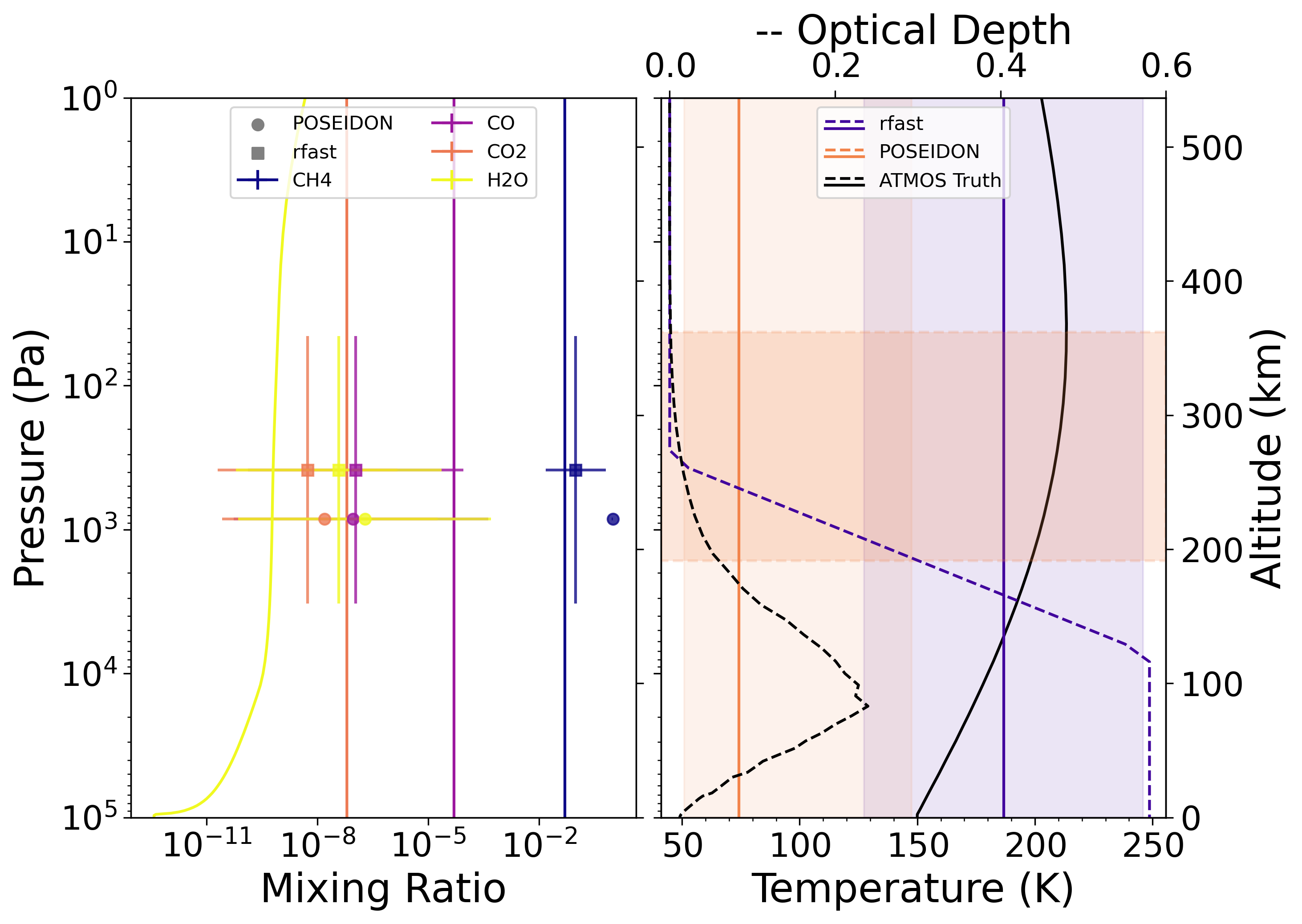}
    \caption{Retrieved temperature, mixing ratios, and haze parameters for the Atmos exo-Titan model. Left: Retrieved volume mixing ratios for all three retrieval cases shown at the approximate depth of the haze layer. Right: Haze opacity as a function of pressure and altitude. The temperature compared to the \Atmos truth are also shown with their errors shown as shaded regions. Note that the \poseidon{} haze location is inferred from the pressure contribution functions.}
    \label{fig:aexohaze}
\end{figure*}

We show the atmospheric components in \autoref{fig:aexohaze} only for the retrieved chemical species, note that the \Atmos model includes the full suite of species mentioned in \autoref{sec:methods}. Despite the similarities in the spectra produced, there are marked deviations in the parameters used to generate that spectrum. First \poseidon retrieves a predominantly methane atmosphere that is then composed of water and the other species to a lesser degree. This is paired with a markedly cold atmosphere and a relatively higher haze layer than is the truth. The \rfast retrieval has a more accurate haze layer and temperature, but also has less of the other species. However, these retrieved temperatures are consistent to within their 1-$\sigma$ confidences.

\section{Discussion}
\label{sec:discussion}

Our results broadly demonstrate the capability of exoplanet atmospheric retrieval models to infer accurate information about a hazy Titan-like atmosphere. Diagnostic spectral features from \ce{CH4} and the hydrocarbon haze, shape the 1--5\,\micron{} transmission spectrum of Titan, while \ce{C3H8}, \ce{C2H2}, and \ce{CO} are also identifiable with higher precision observations. 

\begin{figure}
    \includegraphics[width=\linewidth]{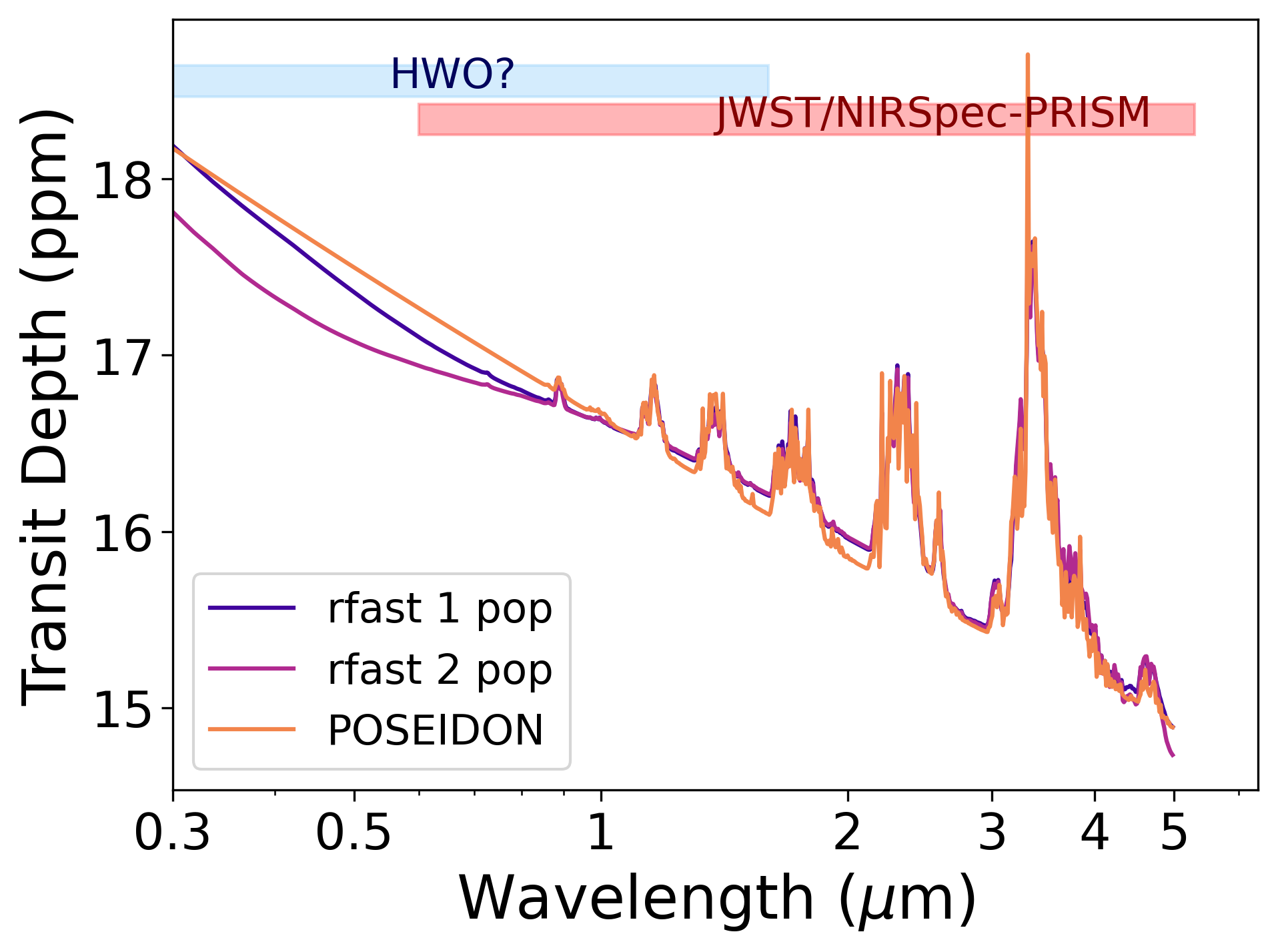}
    \caption{The best fit models from the Cassini Titan scenarios across a wider set of wavelengths encompassing potential HWO coverage. While the spectra are all similar in the 1--5\,\micron{} region, at short wavelengths they begin to diverge.}
    \label{fig:hazediff}
\end{figure}
Even at the precision of the Cassini observation, it is difficult to constrain which of the two haze populations best fits the data. However, a close scrutiny of the particle properties in \autoref{fig:khare} shows that the key difference that particle size has on the spectrum is the slope and the wavelength of its onset. In \autoref{fig:hazediff}, we show the best fit models from the Cassini observation retrieval for all three retrieval scenarios, but extended to shorter wavelengths. We can see how shortward of 1\,\micron{} the spectra all show different slope properties. Additionally data at shorter wavelengths may have been the key in determining which of the \rfast properties was preferred, and a range of particle sizes may yet have been well fit by the simple \poseidon haze slope model. Precise transmission spectra of such hazy Titans at these wavelengths are out of reach for HST, but may yet be possible with HWO \citep{Feinberg2024SPIE} or other future telescopes capable of sub-ppm precisions. 

Previous work by \citet{Changeat2025CloudJupiters}, investigated the information content of spectra taken by JWST with the aim of understanding clouds and hazes in giant planet atmospheres. The use of an average transmission spectrum in that work was intended to better capture the real variations into annulus-averaged exoplanet observations. Previous work by \citet{Niraula2025TheExoplanet} chose only a single Cassini observation however they chose the observations taken at 70\degr{}S, while we chose 27\degr{}N for the ability to compare with \citet{Seignovert2021HazeMission}. In that work they investigated the choice of chemical species on the interpretation of the spectrum using an agnostic power law haze.

In this work, we specifically selected one Cassini observation to explore if the haze sizes and vertical structure can be retrieved. This single observation choice may also be the reason why we did not need to invoke an additional cloud, which \citet{Changeat2025CloudJupiters} showed improved their reproduction of the observations. All three of these studies chose different retrieval tools and our selected species for examination follow the ``base combination" outlined in \citet{Niraula2025TheExoplanet} with the addition of \ce{C2H4}, which is a known species in Titan's atmosphere.

\begin{figure}
    \includegraphics[width=\linewidth]{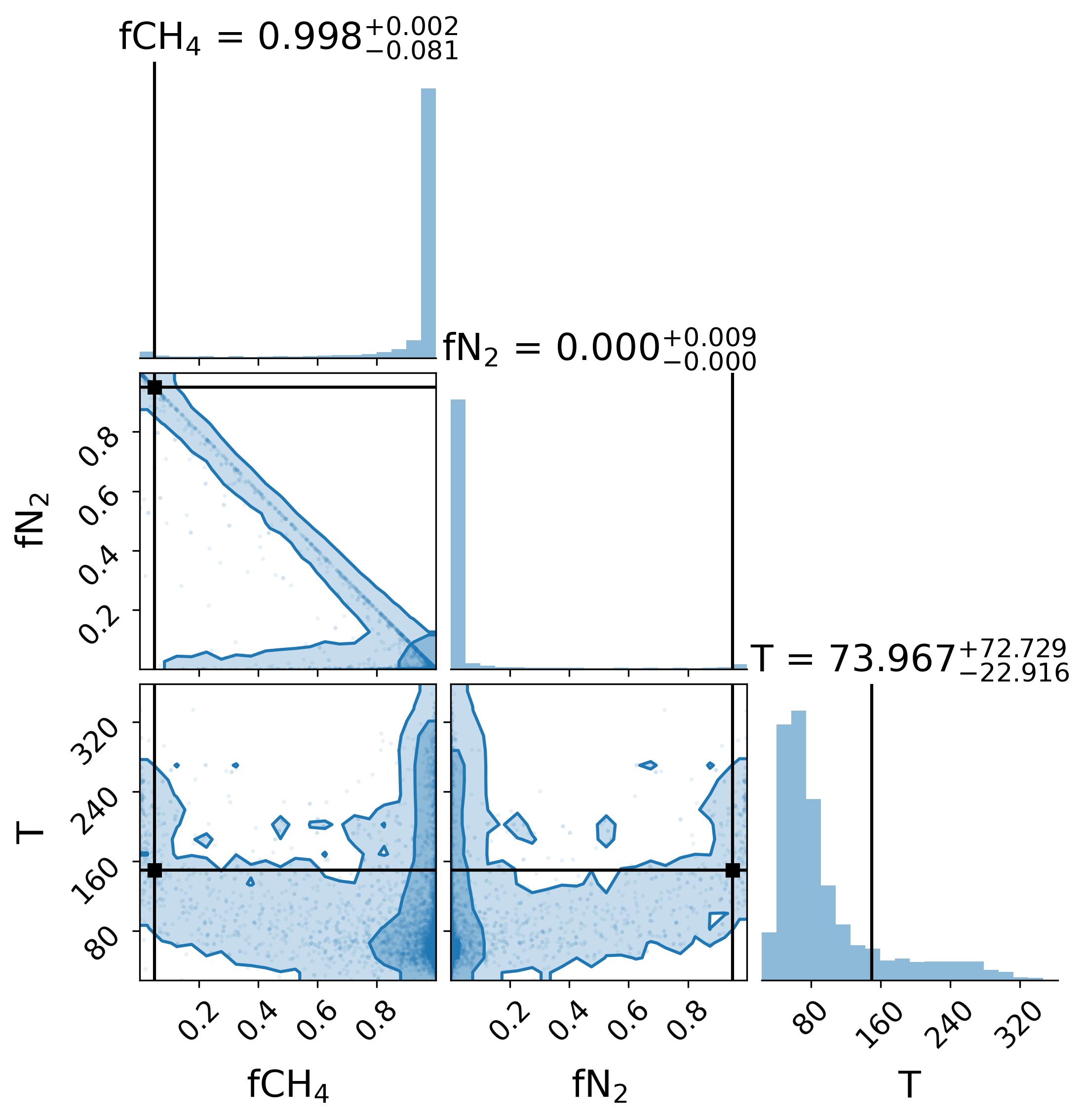}
    \caption{A subset of retrieval posteriors from simulated Atmos exo-Titan, the TRAPPIST-1\,h data for a Titan-like atmosphere obtained using the \poseidon model. The agnostic bulk gas assumption afforded by the CLR prior formalism leads to the incorrect assessment that \ce{CH4} is the bulk gas. However, the true \ce{N2}-dominated Titan atmosphere (black lines and points) does exist within posterior volume along a stark \ce{CH4}-\ce{N2} degeneracy that is mediated by temperature.} 
    \label{fig:poseidon_CLR_corner}
\end{figure}

The use of CLR priors in \poseidon retrievals of JWST quality simulated spectra of TRAPPIST-1\,h reveal a classic challenge intrinsic to small planet atmospheric characterization, that is, ambiguity in the bulk atmospheric constituent. The CLR transformation is used chiefly to encode uncertainty in the identity of the background atmospheric gas \textit{a priori}, so it is agnostic rather than prescribed \citep{Benneke2012AtmosphericSpectroscopy}. \autoref{fig:poseidon_CLR_corner} shows a posterior corner plot for a subset of free parameters, including the \ce{CH4} and \ce{N2} volume mixing ratios (expressed as linear fractions) and the isothermal atmospheric temperature for the \poseidon case presented in \autoref{sec:results:exo}. While this retrieval provides an acceptable fit to the transmission spectrum, the posteriors are biased in favor of an incorrect interpretation where \ce{CH4} is the bulk gas and makes up 99.8\% of the atmosphere by volume with trace amounts of \ce{N2}. This solution is also biased colder than the truth to compensate for the different molecular weights of \ce{CH4} and \ce{N2}. \autoref{fig:poseidon_CLR_corner} reveals a perfectly linear degeneracy between \ce{CH4} and \ce{N2} along which the correct atmosphere is found with \ce{N2} as the bulk gas and warmer atmospheric temperatures. Thus, we observe that a family of \ce{N2}-\ce{CH4} atmospheres are viable interpretations to the observed spectrum,with the correct interpretation notably not fully rejected by the posterior. By allowing each free gas in the retrieval an equal prior probability of making up the bulk of the atmosphere, the CLR priors opened up the possibility of this incorrect interpretation. The preference for \ce{CH4}-dominated is driven by the fact that Titan's atmosphere contains a wealth of \ce{CH4} features, but no strong direct spectroscopic evidence for \ce{N2}, despite including \ce{N2}-\ce{N2} CIA features in the \poseidon retrievals \citep[see e.g.,][]{schwieterman2015detecting}. Therefore, a greater proportion of the posterior volume has methane, as it must to properly fit the data, while the absence of an observational signature of nitrogen limits its necessity in the fit. Thus, while some samples are have the correct \ce{N2}, the strong detection of \ce{CH4} sways the posterior. 
We note that this effect emerged at JWST quality precisions, whereas our CLR retrievals with \poseidon on the high-precision Cassini data correctly obtained bulk \ce{N2}, indicating that signal-to-noise plays an important role in (mis)identifying the background gas. 

This ambiguity is closely related to a well-known limitation of transmission spectroscopy for small planets, wherein the bulk atmospheric composition may not be uniquely constrained. Early theoretical studies showed that molecular features alone cannot always distinguish whether an absorber is a trace gas or the dominant atmospheric constituent \citep{Benneke2012AtmosphericSpectroscopy, Benneke2013HOWSUPER-EARTHS}. This challenge has also been highlighted observationally by the difficulty of distinguishing between \ce{H2}-dominated and high-mean molecular weight (MMW) atmospheres of sub-Neptunes \citep[e.g.,][]{bean2010ground, Kreidbergetal2014flat, wallack2024jwst}. Similar behavior has also been identified in reflected light retrievals, where agnostic priors on the bulk atmospheric constituent can lead to misidentification of the background gas \citep{damiano2025effects}. 

In this context, our Titan retrieval work provides an empirical demonstration, using a Solar System atmosphere with known composition, that agnostic priors on the bulk gas, while notionally advisable, can admit a family of high-MMW solutions in which the dominant atmospheric constituent is misidentified, and a spectrally active gas spuriously favored as the bulk gas. In such cases, care should be taken to report such degeneracies. Furthermore, additional context from self-consistent (photochemical) modeling may be required to break such degeneracies when they are presented in real JWST---or other observatory---spectra where the ground-truth is not known. Thus, our results show how choices in gas mixtures whether they be bulk or trace \citep{Niraula2025TheExoplanet} gas species, can bias retrieval outcomes. These findings highlight the importance of validating exoplanet models and interpretation strategies on high-quality Solar System measurements. 

\section{Conclusions}
\label{sec:concl}
We conducted an examination of the fidelity of two retrieval frameworks in inferring the parameters of Titan and an exo-Titan. We used both actual observations of Titan with Cassini/VIMS-IR and a forward model from \Atmos to baseline retrieval results in both Titan and exo-Titan scenarios. The retrievals varied in their treatment of hazes, where \poseidon uses a rayleigh scattering enhancement model and \rfast uses the prescribed particle properties of \citet{Khare1984OpticalFrequencies} tholins to populate a haze layer of either a singly-sized aggregate or a mix of two-sizes.

Our retrievals on the Titan observations show that both retrieval frameworks are able to reproduce the spectrum with differing, but similar, inferred properties, however once applied to an exo-Titan in a JWST-style observation the resulting parameters are inconsistent despite producing consistent spectra. This shows the importance of the short wavelength capabilities of observatories like HST and in the future, HWO, which can capture the more critical diagnostic wavelength that would trace the differing particle properties of the haze population present in the atmosphere assuming their have the necessary precisions. We did not explore additional effects from haze shapes or alternative compositions, which may also have signatures in this wavelength range.

Our investigation of agnostic bulk atmospheric constituents with \poseidon demonstrate that for moderate-precision spectra expected for rocky planets for the next several decades, agnostic composition priors can lead to misidentification of the bulk atmospheric gas by promoting spectrally active species over spectrally inactive constituents. This further motivates the need to baseline our models and techniques on Solar System objects to understand how our assumptions, whether strong and prescriptive or weak and agnostic, affect our downstream planetary interpretations. Since small planets are out of reach for HST, we look towards future observatories like HWO and what they may be able to do in the nUV -- optical wavelength regime for transiting exoplanets.

\section*{acknowledgments}
This material is based upon work supported by the National Aeronautics and Space Administration (NASA) under Grant Nos. 80NSSC23K0268 and issued through the Exoplanet Research Program (XRP).

\vspace{5mm}
\facilities{NASA Exoplanet Archive, The SAO Astrophysics Data System}

\software{astropy \citep{Astropy}, Atmos \citep{Lincowski2018}, rfast \citep{Robinson2023ExploringObservations}, LBLABC \citep{Meadows1996}, MultiNest/PyMultiNest \citep{feroz_multinest_2009, buchner_x-ray_2014}, \poseidon \citep{MacDonald2017HDWater, MacDonald2023POSEIDON:Spectra}, SMART \citep{Meadows1996}, GNU parallel \citep{Tange2018}, \texttt{PandExo} \citep[v3.0;][]{Batalha2017}}

\bibliography{references,titan}
\bibliographystyle{aasjournal}

\end{document}